\documentclass[12pt,preprint]{aastex}
\makeatletter
\newcommand{\vo}{\vec{o}\@ifnextchar{^}{\,}{}}
\makeatother

\usepackage{amssymb}
\usepackage{amsmath}
\usepackage{graphicx}
\usepackage{subfigure}
\usepackage{latexsym}
\usepackage{xcolor}
\usepackage{float}

\begin{document}

\title{North-south asymmetry in Rieger-type periodicity during solar cycles 19-23}

\author{Eka Gurgenashvili\altaffilmark{1}, Teimuraz V. Zaqarashvili\altaffilmark{2,1,5}, Vasil Kukhianidze\altaffilmark{1},
Ramon Oliver\altaffilmark{3,4}, Jose Luis Ballester\altaffilmark{3,4}, Mausumi Dikpati\altaffilmark{6}, Scott W. McIntosh\altaffilmark{6}}

\altaffiltext{1}{Abastumani Astrophysical Observatory at Ilia State University, Tbilisi, Georgia}
\altaffiltext{2}{IGAM, Institute of Physics, University of Graz, Universit\"atsplatz 5, 8010 Graz, Austria, Email: teimuraz.zaqarashvili@uni-graz.at}
\altaffiltext{3}{Departament de F\'isica, Universitat de les Illes Balears, E-07122, Palma de Mallorca, Spain}
\altaffiltext{4}{Institute of Applied Computing \& Community Code $(IAC^3)$, UIB, Spain}
\altaffiltext{5}{Space Research Institute, Austrian Academy of Sciences,
Schmiedlstrasse 6, 8042 Graz, Austria}
\altaffiltext{6}{High Altitude Observatory, National Center for Atmospheric Research, PO Box 3000, Boulder, Colorado 80307, USA.}

\begin{abstract}
Rieger-type periodicity has been detected in different activity indices over many solar cycles. It was recently shown that the periodicity correlates with solar activity having a shorter period during stronger cycles. Solar activity level is generally asymmetric between northern and southern hemispheres, which could suggest the presence of a similar behavior in the Rieger-type periodicity. We analyse the sunspot area/number and the total magnetic flux data for northern and southern hemispheres during solar cycles 19-23 which had remarkable north-south asymmetry. Using wavelet analysis of sunspot area and number during the north-dominated cycles (19-20) we obtained the periodicity of 160-165 days in the stronger northern hemisphere and 180-190 days in the weaker southern hemisphere. On the other hand, south-dominated cycles (21-23) display the periodicity of 155-160 days in the stronger southern hemisphere and 175-188 days in the weaker northern hemisphere. Therefore, the Rieger-type periodicity has the north-south asymmetry in sunspot area/number data during solar cycles with strong hemispheric asymmetry. We suggest that the periodicity is caused by magnetic Rossby waves in the internal dynamo layer. Using the dispersion relation of magnetic Rossby waves and observed Rieger periodicity we estimated the magnetic field strength in the layer as 45-49 kG in more active hemispheres (north during the cycles 19-20 and south during the cycles 21-23) and 33-40 kG in weaker hemispheres. The estimated difference in the hemispheric field strength is around 10 kG, which provides a challenge for dynamo models. Total magnetic flux data during the cycle 20-23 reveals no clear north-south asymmetry which needs to be explained in the future.
\end{abstract}

\section{Introduction}

Short-term variation in gamma ray flares with period of 155-160 days was discovered by \citet{Rieger1984} during solar cycle 21.  The periodicity later was detected in almost all activity indices \citep{Dennis1985, Bai1987, Lean1989, Bai1990, Lean1990,  Kile1991, Oliver1998, Ballester1999, Krivova2002,  Dimitropolou2008}. \citet{Carbonell1990} and \citet{Carbonell1992} reported the 155-day periodicity in records of the sunspot area during cycles 14-20 and 12-21, respectively. They found that the periodicity was clearly seen during cycles 16-21, but was absent during cycles 12-15. \citet{Ballester2002} analyzed the records of photospheric magnetic flux and found that the periodicity appeared during cycle 21, but it was absent in cycle 22.

The Rieger type periodicity is found also in historical data sets during the earlier cycles. Using two historical aurorae data sets, \citet{Vaquero2010} tried to evaluate presence of Rieger period during the cycles 3-4. They have detected the 150 day period in one auroral dataset during 1777-1781 (cycle 3), but they could not confirm the same periodicity for the cycle 4. \citet{Silverman1990} investigated the occurrence of auroras during 16th and 18th centuries and found 158 and 182-185 days period for the years of 1570-72,1736-39 and 1787-90, respectively.  \citet{Ballester1999} analysed daily number of sunspot groups between 1610 and 1995 and found near 158 day period around the maximum of solar cycle 2. After cycle 2, no strong evidence for the periodicity was found until the 20th century.

Therefore, the Rieger periodicity of 154 days is not a permanent feature of solar activity, but it varies from cycle to cycle. It was also shown that the periodicity usually appears during 1-3 years near the cycle maxima and it may vary from 130 to 185 days \citep{Lean1990, Oliver1998, zaqarashvili2010a}. Recently, \citet{Gurgenashvili2016} analyzed long-term sunspot data for solar cycles 14-24 and showed that the Rieger periodicity is anti-correlated with solar cycle strength: stronger cycles show shorter periods. Observed correlation suggests that the periodicity is related to the dynamo layer in the solar interior.

Most promising explanation of the Rieger-type periodicity is connected to magnetic Rossby waves in the solar tachocline  \citep{zaqarashvili2010a}. The differential rotation and toroidal magnetic field trigger the instability of spherical harmonics of magnetic Rossby waves with period of 155-160 days, which leads to the quasi-periodic emergence of magnetic flux towards the surface. The dispersion relation of magnetic Rossby waves depends on the magnetic field strength \citep{zaqarashvili2007,zaqarashvili2009}, therefore the observed periodicity should depend on solar activity level, which fairly corresponds to observations \citep{Gurgenashvili2016}. Recent discovery of Rossby waves by STEREO and SDO coronal bright point observations \citep{McIntosh2017} fully confirmed the Rossby wave scenario as a mechanism for Rieger-type periodicity.

Solar activity generally shows north-south asymmetry in many indicators  \citep{Sporer1894, Maunder1904, Babcock1959, Waldmeier1971, Roy1977,  Carbonell1993, Oliver1994, Ballester2005,  Temmer2002, Temmer2006, Li2002, Gigolashvili2005, McIntosh2013, McIntosh2014a, McIntosh2014b, McIntosh2015}, which means that the strength of the cycle is different in northern and southern hemispheres. If the Rieger-type periodicity depends on the activity strength, then it should also display the north-south asymmetry. The different periodicity in northern and southern hemispheres then may allow to estimate the difference in magnetic field strength in the dynamo layer over hemispheres, which might be a clue for the understanding of hemispheric asymmetry.

Here we analyze several available hemispheric activity indices in order to find the values of the Rieger periodicity in northern and southern hemispheres separately during activity cycles which have remarkable north-south asymmetry.

\section{North-south asymmetry in solar activity}

We use three different data sets to study the north-south asymmetry in the Rieger-type periodicity:  1) Greenwich Royal Observatory (GRO) daily and monthly sunspot area USAF/NOAA for northern and southern hemispheres  (http://solarscience.msfc.nasa.gov/greenwch.shtml), which are available during 1874-2016, 2) Kanzelh\"ohe Solar Observatory (KSO) and Skalnat\'e Pleso Observatory (SPO) hemispheric sunspot number data (http://vizier.cfa.harvard.edu/viz-bin/VizieR?-source=J/A+A/447/735), which are available in the interval 1945-2004 \citep{Temmer2006}, 3) The Mount Wilson total magnetic flux (MWTF) data which are available between 1966-2002.

North-South asymmetry was also presented during the Maunder minimum (MM, 1645-1715), when the solar activity was extremely low. \citet{Vaquero2015} and \citet{Usoskin2015} analyzed several data sets including both direct and indirect data catalog published by Sp\"orer nearly 130 years ago, sunspot latitudes in the butterfly diagram during MM published by Ribes and Nesme-Ribes almost 20 years ago, aurorae historical reports during MM, Cosmogenic radionuclides etc. They have calculated the asymmetry index using these data sets and confirmed a strong south-dominated hemispherical asymmetry during MM. The Sp\"orer data are given in the paper of \citet{Vaquero2015} and http://haso.unex.es.

We are interested to seek for the Rieger periodicity in the cycles with remarkable north-south asymmetry in order to avoid statistically insignificant correlation between activity and periodicity. Therefore, we first study the long-term north-south asymmetry using GRO sunspot data from 1901 to 2016, which correspond to the cycles 14-24, because earlier data is not fully reliable \citep{Cliver2016,Cliver2017, Erwin2013, Willis2016a, Willis2016b}. Figure 1 (upper panel) shows monthly averaged sunspot area vs time. From coloured polygons one can see that the north-south asymmetry is remarkable near the cycle maxima in most cases and different hemisphere dominates at different phase of corresponding cycle.  For example, the southern hemisphere was more active during the ascending and descending phases of cycle 14, while the northern hemisphere was dominating near the cycle maximum. Similar result was previously noticed by \citet{Newton1955}, who showed that the northern hemisphere was dominant in the early phases of cycles 12 - 15 with a switch to south-dominance later in each cycle. The opposite behaviour was found during cycles 17 - 18. Therefore, full dominance of one hemisphere is not well established.  Cycles 19-23 seem exceptions as the asymmetry in these cycles are very strong and can be considered as statistically significant.

Due to the small value of north-south asymmetry in most cycles, it is very important to study the statistical significance. \citet{Carbonell2007} used several data sets and estimated the statistical significance of north-south asymmetry using different statistical analysis, such as Binomial distribution, Excess, Normal approximation to the Binomial distribution and Pearson's chi-square test. Similar analysis was performed later by \citet{Zhang2015}.

In order to find the statistical significance of north-south asymmetry (SSNSA) in the cycles 19-23 we carried out cycle-to-cycle statistical analysis using the Binomial distribution (see the Table 1)
\begin{equation}
\label{dif1} P_k={{n!} \over {k!(n-k)!}} p^k q^{n-k}.
\end{equation}
where $n$ is the total number of sunspot area, $k$ is the sunspot area for one hemisphere, $p$ is the probability for one hemisphere to be stronger and $q$ is the probability of the another hemisphere. In our case $p=q=0.5$.

\begin{table}[h!]
\centering
 \begin{tabular}{||c c c c c c  ||}

  \hline
 SSNSA \%  & 19  & 20  & 21 & 22 &  23    \\
 \hline\hline
$P < 0.3\%$ & 86 \% & 86 \% & 85 \% & 84.5 \% & 80 \% \\
 \hline
 $0.3\% <P < 5\%$ & 6 \% & 6 \% & 5.7 \% & 5 \% & 5.4 \% \\
 \hline
 $5\%<P<10\%$ & 0.6 \% & 0.6 \% & 0.4 \% & 0.2 \% & 0.2 \% \\
 \hline
 $P>10\%$ & 7.4 \% & 7.5 \% & 8.7 \% & 10.5 \% & 14.6 \% \\
 \hline
\end{tabular}
\caption{Estimated statistical significance of north-south asymmetry during cycles 19-23.}
\label{table:1}
\end{table}

When $P < 0.3\%$, we have a highly significant result, if $0.3\% <P < 5\%$, we have a statistically significant result, if $5\%<P<10\%$, it is marginally significant, and when $P>10\%$, it is a statistically insignificant result. The results in Table 1 show that the level of asymmetry and its statistical significance are high in the cycles 19-23, therefore we use only the data of these cycles for further analysis. Figure 1 shows that the cycles 19-20 are north dominated and cycles 21-23 are south-dominated.



\section{North-south asymmetry in Rieger-type periodicity}


As it is noted in the previous section, we have three data sets: Greenwich observatory daily sunspot area, the joint catalogue of the KSO and SPO, where one can find the daily and monthly sunspot number, as well as smoothed monthly data for both hemispheres separately \citep{Temmer2006} and the Mount Wilson total magnetic flux (MWTF) data (for cycles 20-23, which starts from January 1965 and runs till May 2002).

We used the Morlet wavelet analysis \citep{Torrence1998} to find the Rieger-type timescale in the three data series. Figures 2-3 and 4-5 show the wavelets of north-dominated and south-dominated cycles, respectively. Figure 2 shows the  wavelet analysis performed using GRO data for cycles 19-20. It is clearly seen that the northern hemisphere was dominant in almost whole cycle (panel a). The Rieger-type timescale in cycle 19 was of order of 158-172 days in the northern hemisphere and 172-182 days in the southern hemisphere. The cycle 20 displays the periodicity of 160-165 days in the northern hemisphere and 182-198 days in the southern hemisphere. The cycle-by-cycle global wavelets are computed and plotted alongside each wavelet in sunspot data, where blue (red) color denotes the global wavelet for cycle 19 (20). The global wavelet analysis gives peaks at 160 (180) days in the northern (southern) hemisphere in the cycle 19 and at 165 (190) days in the northern (southern) hemisphere in the cycle 20. Wavelet analysis reveals that the period of the Rieger-type duration is shorter in the northern hemisphere (by 20-25 days) than the southern one during both cycles.

Figure 3 shows the  wavelet analysis of KSO/SPO data during cycles 19-20. The Rieger-type timescale was of order of 158-170 days (with a peak at 165 days) in the northern hemisphere and 174-190 (with a peak at 175 days) days in the southern hemisphere in cycle 19. In the cycle 20, the Rieger periodicity was 151-156 days (with a peak at 155 days) in the northern hemisphere and 185-190 days (with a peak at 188 days) in the southern hemisphere. KSO/SPO data also show that the stronger northern hemisphere displays shorter periodicity than the weaker southern hemisphere. Hence the north-south behavior of the Rieger periodicity agree qualitatively in GRO and KSO/SPO in the cycles 19 and 20.

The N-S asymmetry in the cycles 21-23 shifted to the southern hemisphere \citep{Verma1992}. We performed the wavelet analysis of the south dominated cycles separately for sunspot data. Figure 4 represents the wavelet analysis of GRO data for the south-dominated cycles 21-23. The global wavelets are plotted on right-hand-side, where blue, black and red colors correspond to the cycle 21, 22 and 23, respectively. As it is expected, the weaker northern hemisphere now shows longer periodicity: 160-187 days with peak at 183 days for the cycle 21, 168-190 days with peak at 180 days for the cycle 22 and 170-185 days (peak at 175 days) in the cycle 23. The stronger southern hemisphere displays the shorter periodicity of 155-165 days with peak at 158, 160 and 160 days, for the cycles 21-23, respectively (see the table 2).  The difference between hemispheric periodicity is around 15-23 days very similar to the north-dominated cycles.

Figure 5 shows the wavelet analysis of KSO/SPO data for south dominated cycles 21-23. The periodicity in northern hemisphere is of the order of 180-190 days (peak at 188 days) in cycle 21, 175-190 days (peak at 177 days) in cycle 22 and 165-185 days (peak at 174 days) during cycle 23. Stronger southern hemisphere shows the period of 150-165 days with peaks at 155, 158 and 161 days for the cycles 21, 22 and 23, respectively. However, the cycle 22 displays another stronger peaks at 190 days in the southern hemisphere in both GRO and KSO/SPO data, which is out of general picture in N-S asymmetry. This interesting disagreement will be discussed later.

Figure 6 presents MWTF data during cycles 20-23 with corresponding wavelet analysis. The upper panel shows that only cycle 22 displays remarkable N-S asymmetry with more active southern hemisphere. The cycles 20, 21 and 23 have almost no hemispheric asymmetry. Wavelet analysis gives the periodicity of 160-172 days (with a peak at 168 days) in the cycle 20, 160-180 days (with peak at 170 days) in the cycle 21, 165-180 day period (peak at 175) in the cycle 22 and 160-175 days (with a peak at 170 days) in cycle 23 in the northern hemisphere. The southern hemisphere shows the periodicity of 158-168 days (with a peak at 165 days) in the cycle 20, 180-190 days (with a peak at 187 days) in the cycle 21, 150-160 days (with a peak at 155 days) in the cycle 22 and 160-180 days (with a peak at 170 days) in cycle 23. In contrast with GRO and KSO/SPO data, the total magnetic flux shows no clear north-south asymmetry in the Rieger periodicity during cycles 20 and 23. The cycles 21-22 show some N-S asymmetry in magnetic flux but not as significant as in the sunspot data.


The wavelet analysis of sunspot data (GRO, KSO/SPO) clearly show that the Rieger timescale is characterized by the hemispheric asymmetry: the stronger hemisphere displays shorter periodicity of the order of 160-165 days, while weaker hemisphere displays longer periodicity of the order of 175-190 days. This result fairly agrees with the finding of \citet{Gurgenashvili2016} that the stronger cycles generally show shorter periodicity. Here the hemisphere (e.g. northern hemisphere in cycles 19-20 and southern hemisphere in cycles 21-23) with higher activity level has shorter periodicity.

In addition, activity maxima during cycles 19-20 are shifted with 1-2 years in northern and southern hemispheres (see Figures 2a and 3a). The southern hemisphere reaches its maximum before the northern hemisphere during cycle 19, while it is opposite during the cycle 20 where northern hemisphere reaches the maximum first. The north-south phase shift of solar cycles in sunspot data was studied in details by \citet{Dikpati2007}. They showed that the shift of cycle maxima is more pronounced than the shift of minima (see Figure 5 of the paper). Our result fairly agrees with their finding. The Rieger periodicity displays the similar phase shift as it is seen on Figures 2 and 3. This is in agreement with the previous result that the Rieger periodicity in full disc data appears near the cycle maxima.

On the other hand, the Rieger periodicity shows different behavior in the total magnetic flux. Here no clear north-south asymmetry is seen. \citet{Howard1974} examined magnetic flux data from Mount Wilson magnetograph during 1967-1973 and reported that the total flux in the north was greater than in the south by only a 7\%, therefore asymmetry is missing in the MWTF data. \citet{Chumak2003} studied the behavior of the total sunspot area and magnetic flux during the year 1989 and showed that there is not always positive correlation between active regions and total magnetic flux: sometimes the flux increases or decreases, while the sunspot areas remain the same. The difference between the Rieger periodicity in sunspot area/number (GRO, KSO/SPO) and total magnetic flux (MWTF) can be related with the lack of the permanent positive correlation. The lack of correlation might reflect the fact that the total magnetic flux is a sum of strong sunspot and weak plage fluxes which may have different behavior. During cycle 21, \citet{Rabin1991} found quasi periodic pulsations only in the strong flux, which were uncorrelated between the hemispheres until 1983, than they appear to be synchronized. \citet{Ballester2002} studied MWTF data for cycles 20-23 and found a correlation between impulses in strong flux and flares, but not with weak flux. On the other hand, \citet{Lean1989} reported that the Rieger periodicity was not significant in the plage index. This point surely needs more detailed study.

\section{Discussion}

Rieger type periodicity has been detected during last two centuries in different activity indices, which showed that it is not a permanent feature of the solar activity but varies from cycle to cycle. It was recently shown that the Rieger periodicity correlates with solar cycle strength being shorter during stronger cycles and therefore it could be related to the internal dynamo layer, where strong toroidal magnetic field is generated \citep{Gurgenashvili2016}. Quasi-periodic variation of the dynamo magnetic field with Rieger-type periodicity triggers corresponding variations in activity indices owing to the modulation of erupted magnetic flux. If the Rieger periodicity is the feature of the dynamo layer then it may carry information about its physical parameters.

The mechanism of solar activity still remains as one of the major unsolved problems in solar physics, but the cycles are supposed to be caused by large-scale dynamo action in the solar interior \citep{Charbonneau2010}. The tachocline, thin layer between radiative and convective envelopes, is suggested to be the location of dynamo action. However, there are also dynamo models without tachocline. The magnetic field strength according to the dynamo models without tachocline is less than 10 kG, but the models with tachocline predict much stronger field ($>$ 10 kG) \citep{Charbonneau2013}. Therefore, the estimation of the magnetic field strength is very important as it may put some limitation on dynamo models in the solar/stellar interiors.

Solar activity displays different levels of activity between northern and southern hemispheres. This north-south asymmetry is generally small with weak statistical significance, but it becomes remarkable during some (more stronger) cycles. The asymmetry probably reflects the difference between dynamo magnetic field strengths in northern and southern hemispheres, but the mechanism of the difference is unknown. Even rough estimation of the difference between hemispheric magnetic fields in the dynamo layer may give us a hint to understand the triggering mechanism for the asymmetry. The strength of dynamo magnetic field in different hemispheres can be estimated from the observed Rieger periodicity in hemispheric data.

We used the hemispheric data of GRO daily and monthly sunspot area, joint KSO/SPO daily and monthly sunspot numbers and the Mount Wilson total magnetic flux to find the Rieger periodicity in northern and southern hemispheres during cycles 19-23, when the north-south asymmetry of solar activity was remarkable (see Figure 1, upper panel). Figure 1 shows that the northern hemisphere was much more active during the cycles 19-20, but the southern hemisphere became stronger during the cycles 21-23. Wavelet analysis of sunspot data (GRO, KSO/SPO) revealed that the Rieger periodicity was significantly different in both hemispheres being 160-165 days in the northern hemisphere and 175-190 days in the southern hemisphere during north-dominated cycles, while it became 155-160 days in the northern hemisphere and 175-188 days in the southern hemisphere during the south-dominated cycles (see Table 2 for details). Therefore, the periodicity clearly reflects the north-south asymmetry in solar activity.

\begin{table}[h!]
\centering
 \begin{tabular}{||c c c c c c c  ||}

  \hline
 Cycle  & Period  & Period  & Period & Period &  Period &  Period   \\
 Number &( N, GRO) & (S, GRO) & (N, KSO/SPO)& (S,KSO/SPO)& (N, MWTF) & (S, MWTF)  \\ [0.5ex]
 \hline\hline
 19 & 158 & 177 & 156 & 176 & - & - \\
 \hline
 20 & 165 & 190 & 152& 188 & 168 & 165 \\
 \hline
 21 & 183 & 158 & 188& 155 & 170 & 187 \\
 \hline
 22 & 180 & 160 & 177& 158 & 175 & 155 \\
 \hline
 23 & 175 & 160 & 174& 161 & 170 & 170 \\
 \hline

\end{tabular}
\caption{Estimated Rieger Periods (days) for both hemispheres from GRO (column $2-3$), KSO/SPO (column $4-5$) and MWTF Data (column $6-7$), for Solar Cycles 19-23.}
\label{table:1}
\end{table}

\citet{Sturrock1999} suggested that the Rieger periodicity might be caused by r-modes of rotating Sun, which are hydrodynamic (HD) Rossby waves. Then, \citet{Lou2000} suggested an explanation for the periodicity in terms of equatorially trapped HD Rossby waves. However, the periodicity is usually observed in activity indices, hence the magnetic field should be clearly involved in the scenario. Zaqarashvili et al. (2010a) showed that the Rieger periodicity is related to the instability of magnetic Rossby waves due to the differential rotation and toroidal magnetic field in the dynamo layer.
Therefore, the observed periodicity alongside with the dispersion relation of magnetic Rossby waves could lead to the estimation of dynamo magnetic field in individual cycles. Based on the magnetic Rossby wave theory, \citet{Gurgenashvili2016} estimated the magnetic field strength in the dynamo layer being $\approx$ 40 kG during stronger solar cycles (16-23) and $\approx$ 20 kG during weaker cycles (14-15 and 24).

The dispersion relation of fast magnetic Rossby waves (the slow magnetic Rossby waves may lead to the long-term variation of solar cycles as suggested by \citet{zaqarashvili2015}) in the dynamo layer can be written as \citep{Gurgenashvili2016}
\begin{equation}
\label{disp_f} \omega_f={-m \Omega_0}{{1+s_2 + \sqrt{(1+s_2)^2+{{4 B^2_{max}} \over {4 \pi \rho \Omega^2_0 R^2_0 }}{n(n+1)}}}\over{n(n+1)}},
\end{equation}
where $\omega_f$ is the frequency of fast magnetic Rossby waves, $\Omega_0$ is the equatorial angular velocity, $s_2$ is the parameter of the differential rotation, $\rho$ is the density, $R_0$ is the distance from the solar center to the dynamo layer, $B_{max}$ is the dynamo magnetic field strength at 45 degree, $m$ and $n$ are toroidal and poloidal wave numbers, respectively. Only the magnetic field strength is unknown parameter in the dispersion relation, therefore it can be deduced from the observed periodicity. \citet{Gurgenashvili2016} showed that the spherical harmonic with $m$ = 1 and $n$ = 4 may confidently explain the observed periodicity for $30-50\;$ kG magnetic field.

We use the dispersion relation (Eq. 2) for estimation of magnetic field strength in the northern and southern hemispheres during cycles 19-23. The differential rotation parameters were not estimated for the northern and southern hemispheres separately for these cycles, therefore initially we set $s_2=0$ in the equation (2). Based on the GRO data, we calculate the maximum magnetic field strength as 48 kG (38 kG) in the northern (southern) hemisphere during north-dominated cycle 19, 45 kG (33 kG) in the northern (southern) hemisphere during north-dominated cycle 20, 49 kG (36 kG) in the southern (northern) hemisphere during south-dominated cycle 21, 48 kG (38 kG) in the southern (northern) hemisphere during south-dominated cycle 22 and 48 kG (40 kG) in the southern (northern) hemisphere during south-dominated cycle 23. These calculations show that the difference between dynamo magnetic field strengths in northern and southern hemispheres during cycles 19-23 is of the order of 10 kG, which is a quite significant value (see Figure 7). Non-zero differential rotation parameter $s_2$ in Eq. (2) changes the estimated value of magnetic field strength (see the Table 3), however the hemispheric difference still remains of the order of 10 kG. It must be mentioned, however, that the estimation of magnetic field strength is rather rough and future detailed analysis (including numerical simulations) is needed to increase the accuracy. Figure 7 shows that the estimated magnetic field strength does not significantly vary during cycles 21-23, while the cycle amplitude has been continuously declining. This may support the evidence that the sunspot cycle is an "interference" pattern of overlapping 22-year bands \citep{McIntosh2014a}. Moreover, it is seen from Figure 7 that the difference between southern and northern hemispheric magnetic field strengths is also decreasing, which could be a result of interaction of the bands. This point needs detailed study in the future.

\begin{table}[h!]
\centering
 \begin{tabular}{||c c c c c c c c c c c ||}

  \hline
Cycle number  & 19 &19  & 20 &20 & 21 & 21 & 22 &22 &  23 &23 \\
Differential rotation, $s_2$   &$0.19$ & 0 &$0.16$ & 0 & $0.14$ & 0 & $0.14$ & 0 & $0.17$ & 0 \\ [0.5ex]
 \hline\hline
 $B_{max}$ (kG), north & 40 & 49 & 37 & 45 & 28 & 36 & 30 & 38 & 31 & 40 \\
 \hline
$B_{max}$ (kG), south & 28 & 39 & 23 & 33 & 43& 49 & 42 & 48 & 40 & 48 \\
 \hline

\end{tabular}
\caption{Estimated Magnetic field strength for northern and southern hemispheres during the cycles 19-23. The meanings of differential rotation parameter $s_2$ are obtained by \citet{Javaraiah2005}.}
\label{table:1}
\end{table}

The estimated large difference between dynamo field strengths in the two hemispheres needs to be explained in the future. It may become as a key point to resolve the problem of solar dynamo and activity cycles. It is possible that the observed north-south asymmetry is owing to the overlapping of 11-year oscillating dynamo magnetic field with some steady field component. In this case, the steady field of 5 kG may cause required 10 kG difference in hemispheric magnetic field. \citet{Dikpati2006} showed that the steady (non-reversing) toroidal field can be generated in the lower tachocline due to a steady dynamo in the case of low magnetic diffusivity with the strength of $>$ 1 kG, which is in the range of required value. Then the temporal variation of the non-reversing magnetic field with longer time scales caused by slow magnetic Rossby waves below the solar tachocline \citep{zaqarashvili2015} may lead to the observed variations in north-south asymmetry. This is, however, only speculation and no real physical mechanism resolving the north-south asymmetry problem exists up to now. Recent flux transport dynamo simulations have addressed this problem in terms of N/S asymmetries in surface poloidal source \citep{Belucz2013a} and in meridional circulation \citep{Belucz2013b}, but reason of such asymmetries in the dynamo ingredients is yet to be physically explored.

It must be noted here that the sunspot number data in the cycle 22 displays the significant peak at longer period ($\sim$ 190 days) in the southern hemisphere, which is somehow out of regularity. This long-period peak may correspond to the higher harmonic of magnetic Rossby waves. For example, if the shorter period of 158 days is caused by m=1, n=4 harmonic (as it is suggested above) then the harmonic with m=1, n=5 would give the period of $\sim$ 210 days, which is not far from the observed peak. The long period peaks can be seen also in other cycles and might correspond to the regular pattern. It requires further detailed study.

In contrast of sunspot number/area data, total magnetic flux does not show any remarkable north-south asymmetry in the Rieger periodicity. Therefore, it seems that the total magnetic flux does
not clearly manifest the north-south asymmetry. This is probably caused by the fact that used MWTF contains both, strong sunspot flux and weak plage flux, from which only the strong flux
has N-S asymmetry. This is an interesting question to be answered in the future.

\section{Conclusions}

We carried out the wavelet analysis of the hemispheric sunspot area (GRO), sunspot number (KSO/SPO) and Mount Wilson Total Magnetic flux data during solar cycles (19-23) with remarkable north-south asymmetry: the northern hemisphere was dominated during cycles 19-20 and the southern one was dominated during the cycles 21-23. The analysis of sunspot area/number data showed that the Rieger type periodicity is also asymmetric with hemispheres. We obtained the periods of 160-165 days in the northern hemisphere and 180-190 days in the southern hemisphere during cycles 19-20, while 155-160 days in the northern hemisphere and 175-188 days in the southern hemisphere during the cycles 21-23. Therefore, the Rieger-type periodicity in sunspot area/number data correlates with hemispheric activity levels in the same sense as it correlates with cycle strength based on full disc data \citep{Gurgenashvili2016}: the hemisphere with stronger activity displays the periodicity with shorter period. Hence, the Rieger periodicity is connected to the internal dynamo layer, where the magnetic field and the solar cycles are generated. The magnetic field might be modulated by magnetic Rossby waves, which leads to the quasi-periodic emergence of magnetic flux. This scenario is fully supported by recent direct observations of Rossby waves using STEREO and SDO coronal bright point data \citep{McIntosh2017}. In addition, activity manifests a phase shift of 1-2 years between northern and southern hemispheres, which is clearly seen during the cycles 19-20(see more detailed analysis in \citet{Dikpati2007}). The Rieger periodicity also takes place at different times (with similar 1-2 year shift) in the two hemispheres which means that the quasi-periodic flux emergence correlates to the maximum phase of solar cycles. The obtained periodicity and the dispersion relation of magnetic Rossby waves were used to estimate the magnetic field strength in the tachocline as 45-48 kG in more active hemisphere (northern hemisphere during the cycles 19-20 and the southern one during cycles 21-23) and 32-38 kG in the weaker hemisphere. The north-south difference in the dynamo magnetic field strength is almost 10 kG, which reaches to almost 25 \% of estimated magnetic field. The significant hemispheric difference of the field strength induces future challenges for dynamo models.

\begin{figure}
\includegraphics[width=\columnwidth]{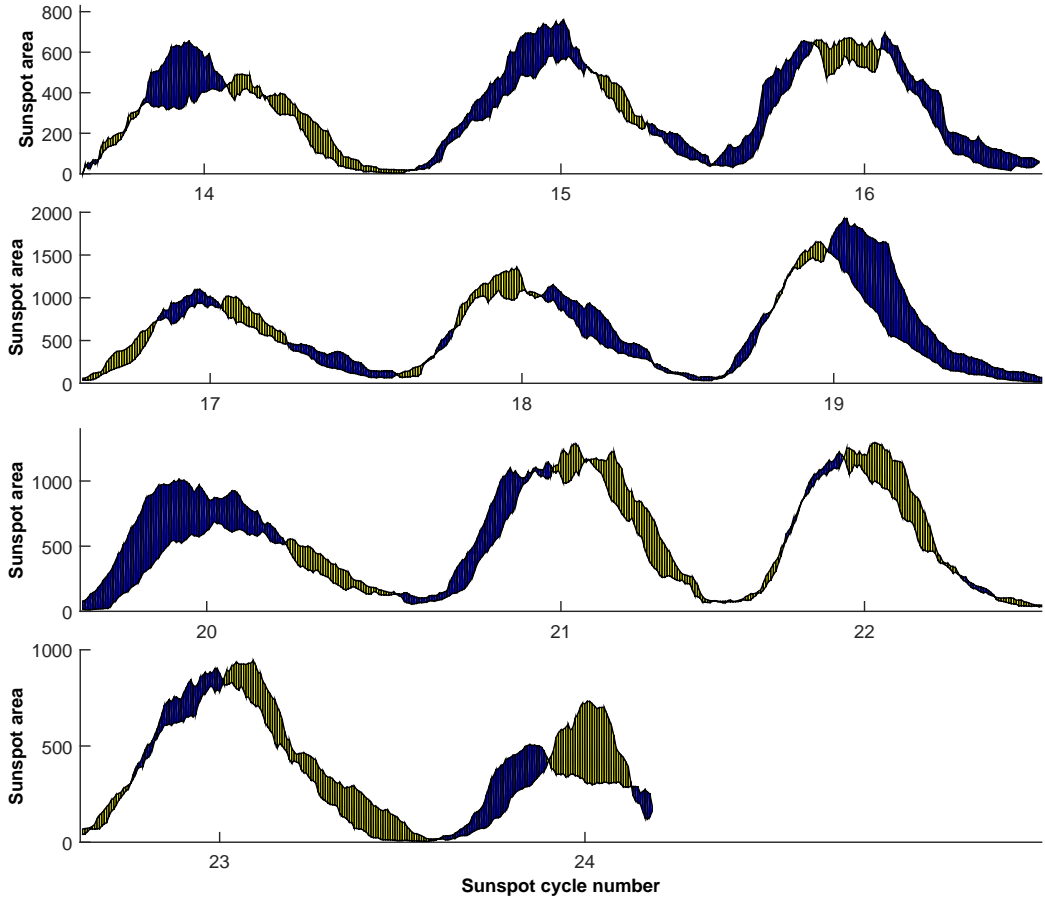}
\caption{Greenwich Royal Observatory monthly averaged hemispheric sunspot area data for cycles 14-24. Blue color denotes excess of the northern hemisphere, and yellow shades correspond to the excess of the southern hemisphere.}
\end{figure}

\begin{figure}
\includegraphics[width=\columnwidth]{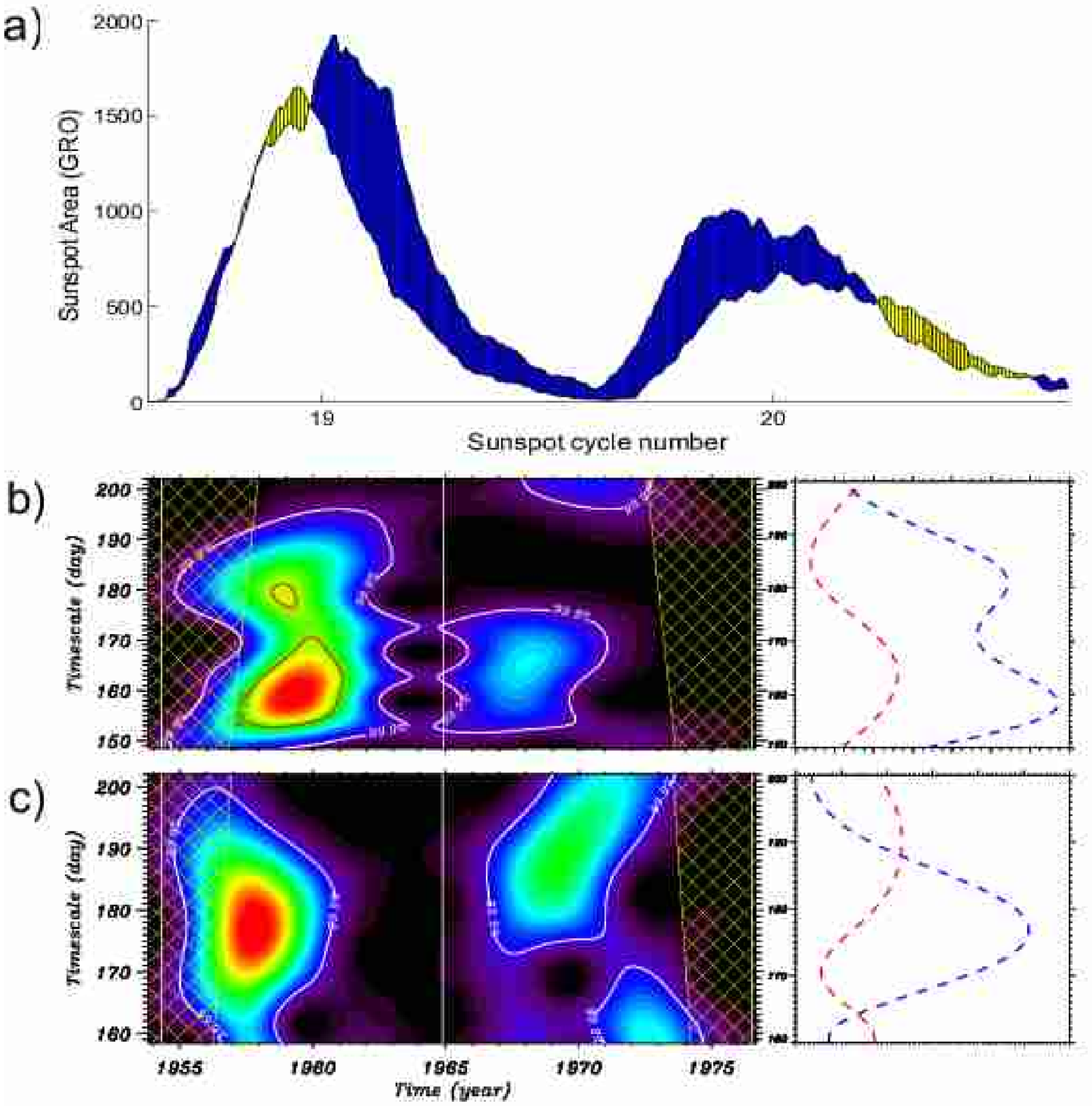}
\caption{Top panel (a) represents GRO monthly averaged hemispheric sunspot area for cycles 19-20 with blue (yellow) color in case of excess northern (southern) hemisphere. Middle panel (b) and bottom panel (c) represent the wavelet of northern and southern hemisphere, respectively. Global wavelet results are plotted to the right, where blue (red) color corresponds to cycle 19 (20).}
\end{figure}

\begin{figure}
\includegraphics[width=\columnwidth]{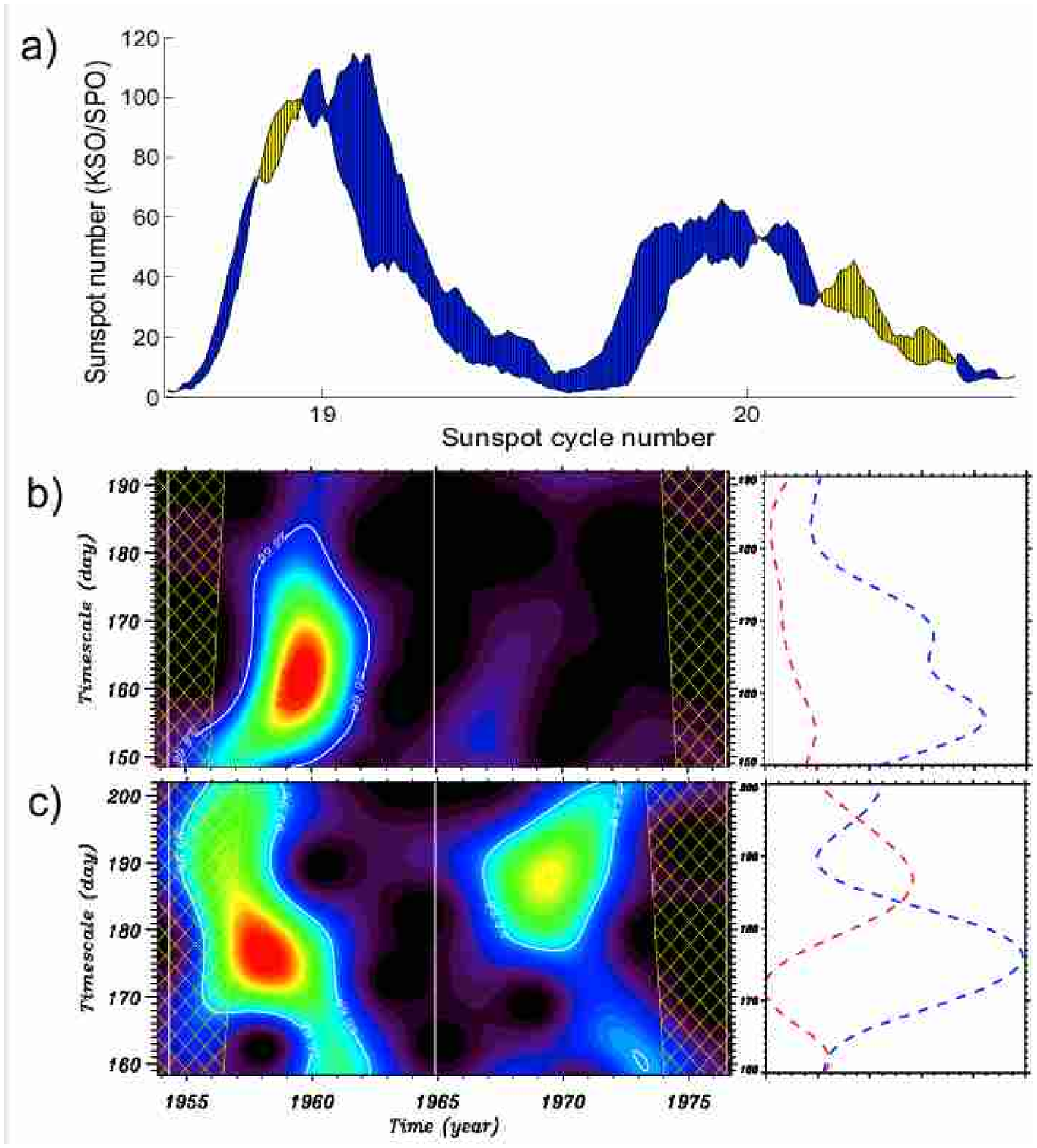}
\caption{Top panel (a) represents KSO/SPO monthly averaged hemispheric sunspot number for cycles 19-20 with blue (yellow) color in case of excess northern (southern) hemisphere. Middle panel (b) and bottom panel (c) represent the wavelet of northern and southern hemispheres, respectively. Global wavelet results are plotted to the right, where blue (red) color corresponds to the cycle 19 (20).}
\end{figure}

\begin{figure}
\includegraphics[width=\columnwidth]{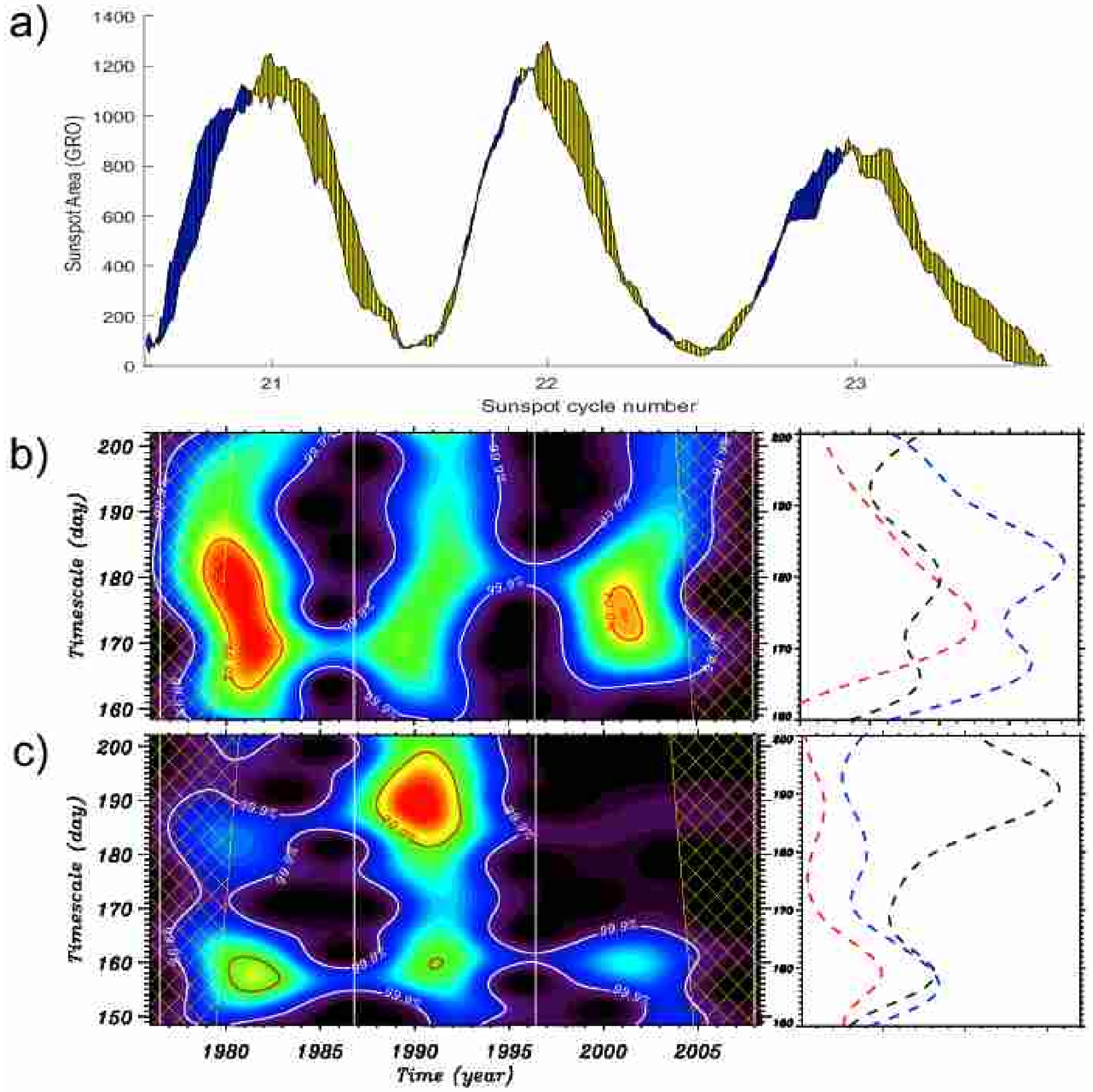}
\caption{Top panel (a) represents GRO monthly averaged hemispheric sunspot area for cycles 21-23 with blue (yellow) color in case of excess northern (southern) hemisphere. Middle panel (b) and bottom panel (c) represent the wavelet of northern and southern hemispheres, respectively. Global wavelet results are plotted to the right, where blue, black and red colors correspond to the cycles 21, 22 and 23, respectively.}
\end{figure}

\begin{figure}
\includegraphics[width=\columnwidth]{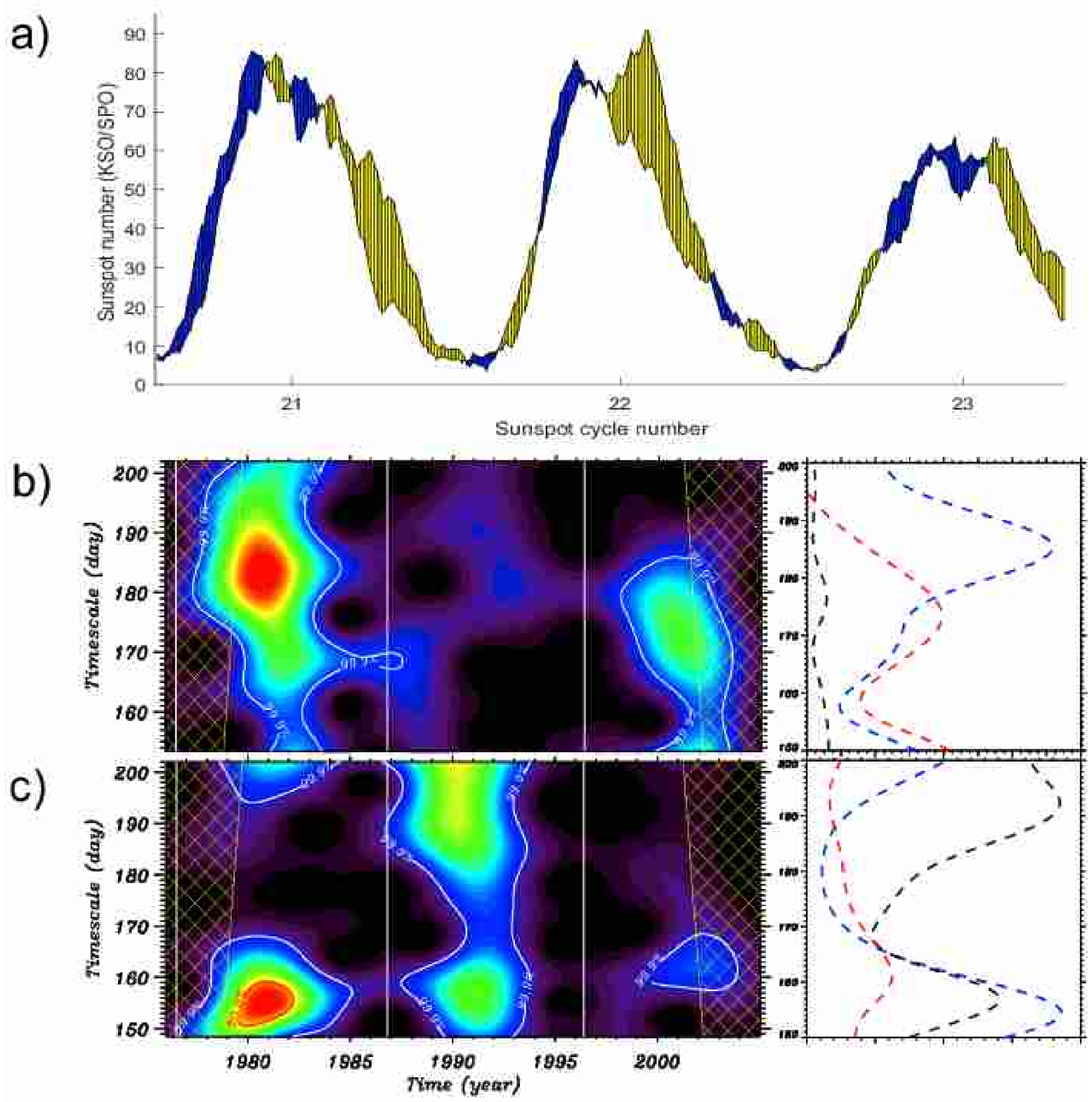}
\caption{Top panel (a) represents KSO/SPO monthly averaged hemispheric sunspot number for cycles 21-23 with blue (yellow) color in case of excess northern (southern) hemisphere. Middle panel (b) and bottom panel (c) represent the wavelet of northern and southern hemispheres, respectively. Global wavelet results are plotted to the right, where blue, black and red colors correspond to the cycles 21, 22 and 23, respectively.}
\end{figure}

\begin{figure}
\includegraphics[width=\columnwidth]{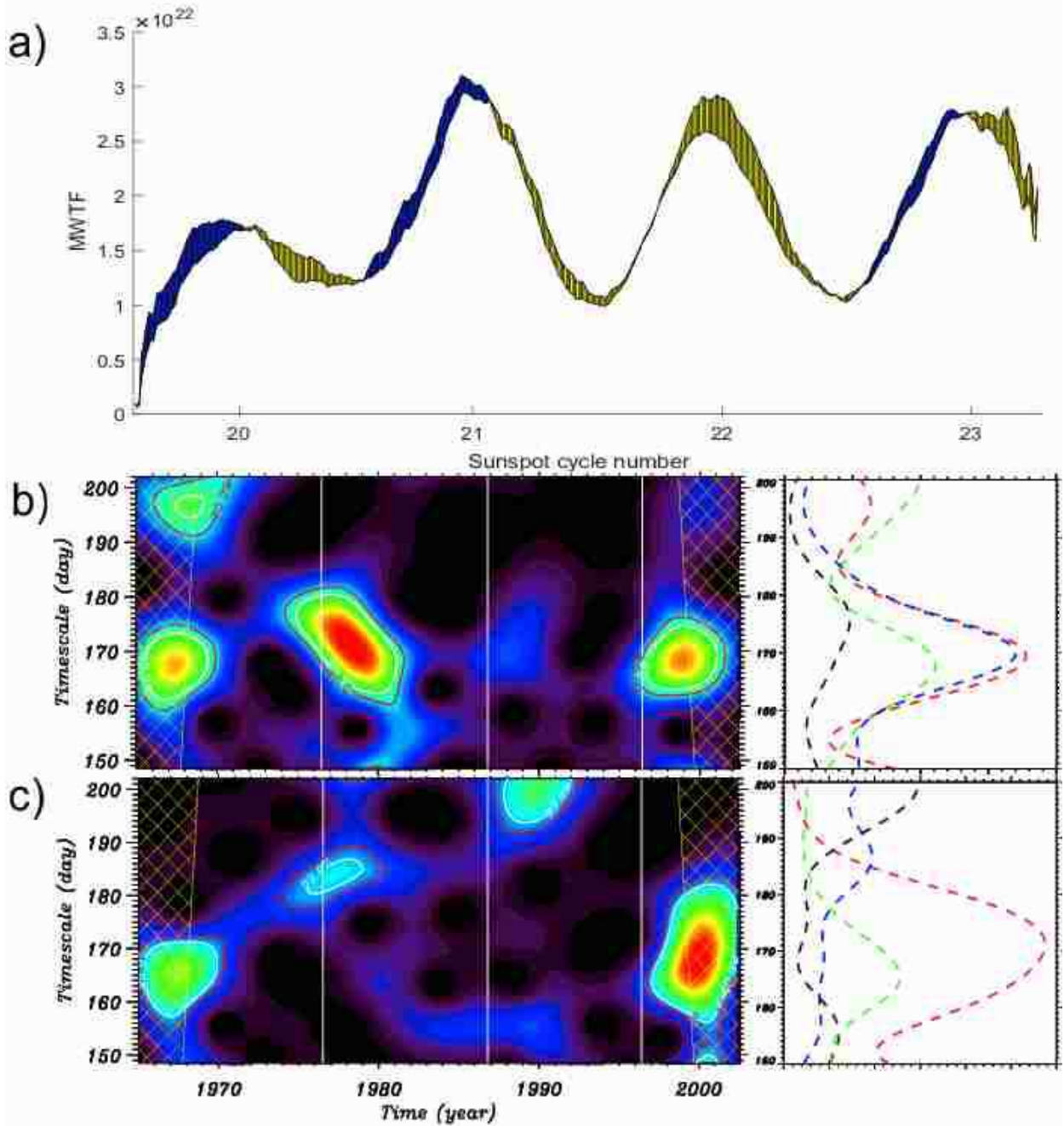}
\caption{Top panel (a) represents Mount Wilson total magnetic flux daily hemispheric data for cycles 20-23 with blue (yellow) color in case of excess northern (southern) hemisphere. Bottom (middle) left panel represent the wavelet of northern (southern) hemisphere. Global wavelet results are plotted to the right, where green, blue, black and red lines correspond to cycle 20-23, respectively.}
\end{figure}

\begin{figure}
\includegraphics[width=\columnwidth]{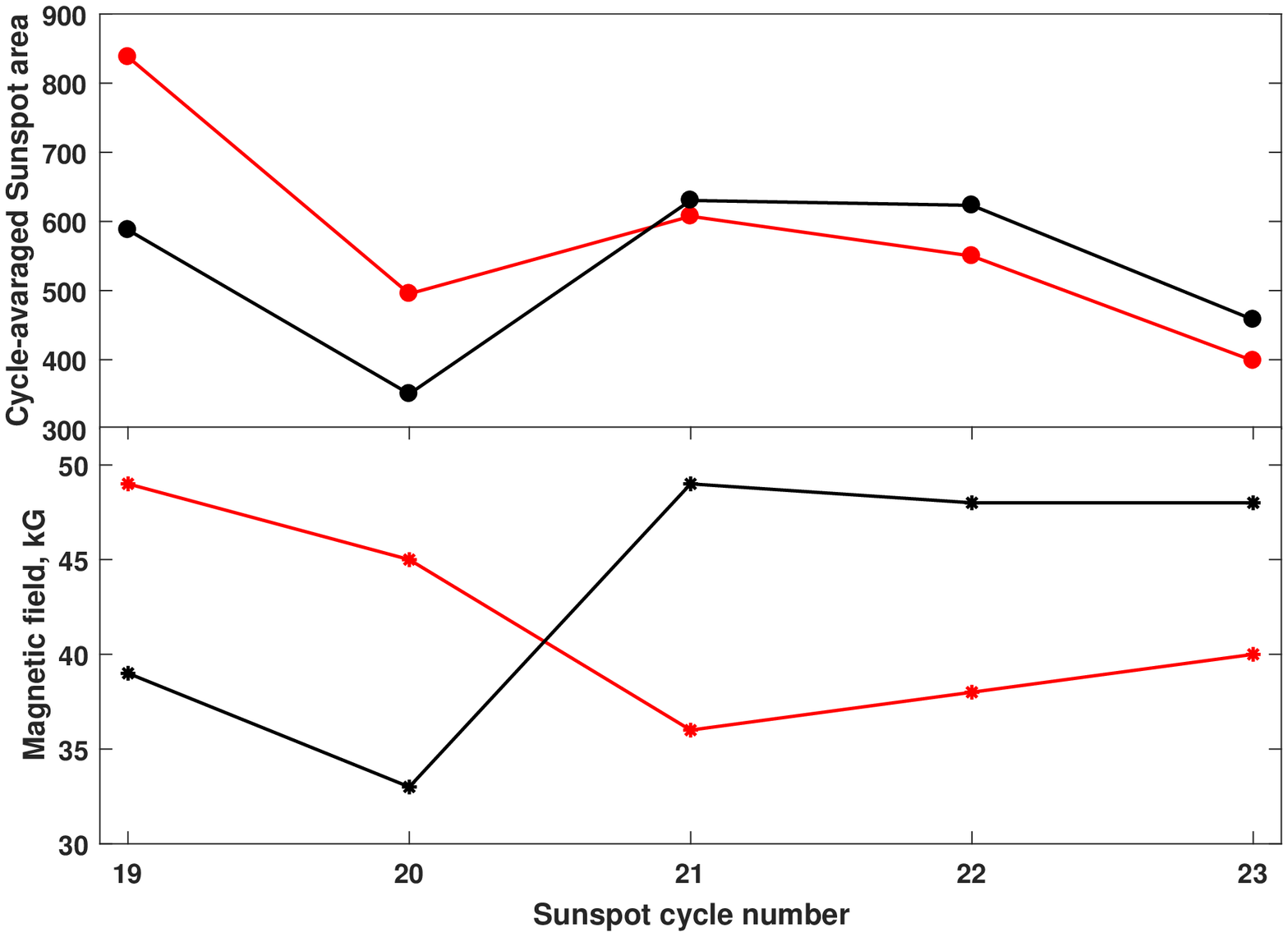}
\caption{Upper panel: cycle-averaged sunspot area for the cycles 19-23 (based on GRO data), where red (black) line corresponds to the northern (southern) hemisphere. Lower panel: estimated magnetic field strength in  northern (red) and southern (black) hemispheres for the cycles 19-23.}
\end{figure}

{\bf Acknowledgements} We thank J. Boyden  who  kindly  provided  us  with  the MWTF  records. The Mount  Wilson  150  Foot  Solar  Tower  is  operated  by UCLA, with funding from NASA, ONR, and NSF, under agreement with the Mount Wilson Institute. This work was supported by Georgian Shota Rustaveli National Science Foundation (projects PhDF2016-130 and 217146) and by the Austrian ``Fonds zur F\"{o}rderung der wissenschaftlichen Forschung'' (FWF) project P26181-N27.  This paper is resulted from discussions at the workshop of ISSI (International Space Science Institute) team (ID 389) "Rossby waves in astrophysics" organized in Bern (Switzerland).

\clearpage

\end{document}